\documentclass[sigconf]{acmart}

\AtBeginDocument{%
  }

\setcopyright{acmlicensed}
\copyrightyear{2025}
\acmYear{2025}
\acmDOI{10.1145/XXXXXXX.XXXXXXX}

\acmISBN{978-1-4503-XXXX-X/25/11}

\usepackage{booktabs}
\usepackage{multirow} 
\usepackage{makecell}
\usepackage{float}
\usepackage{xcolor}

\begin{document}

\title{3S-Trader: A Multi-LLM Framework for Adaptive Stock Scoring, Strategy, and Selection in  Portfolio Optimization}


\author{Kefan Chen}
\affiliation{%
  \institution{The University of Adelaide, Australia}
  \country{}
  }
\email{kefan.chen@student.adelaide.edu.au}

\author{Hussain Ahmad}
\affiliation{%
  \institution{The University of Adelaide, Australia}
  \country{}
  }
\email{hussain.ahmad@adelaide.edu.au}

\author{Diksha Goel}
\affiliation{%
  \institution{CSIRO's Data61, Australia}
  \country{}
  }
\email{diksha.goel@csiro.au}

\author{Claudia Szabo}
\affiliation{%
  \institution{The University of Adelaide, Australia}
  \country{}
  }
\email{claudia.szabo@adelaide.edu.au}

\begin{abstract}
Large Language Models (LLMs) have recently gained popularity in stock trading for their ability to process multimodal financial data. However, most existing methods focus on single-stock trading and lack the capacity to reason over multiple candidates for portfolio construction. Moreover, they typically lack the flexibility to revise their strategies in response to market shifts, limiting their adaptability in real-world trading. To address these challenges, we propose \textbf{3S-Trader}, a training-free framework that incorporates scoring, strategy, and selection modules for stock portfolio construction. The scoring module summarizes each stock’s recent signals into a concise report covering multiple scoring dimensions, enabling efficient comparison across candidates. The strategy module analyzes historical strategies and overall market conditions to iteratively generate an optimized selection strategy. Based on this strategy, the selection module identifies and assembles a portfolio by choosing stocks with higher scores in relevant dimensions. We evaluate our framework across four distinct stock universes, including the Dow Jones Industrial Average (DJIA) constituents and three sector-specific stock sets. Compared with existing multi-LLM frameworks and time-series-based baselines, 3S-Trader achieves the highest accumulated return of \textbf{131.83\%} on DJIA constituents with a Sharpe ratio of \textbf{0.31} and Calmar ratio of \textbf{11.84}, while also delivering consistently strong results across other sectors.
\end{abstract}




\maketitle
\settopmatter{printfolios=true}

\begin{figure*}[h]
\includegraphics[width=0.8\textwidth]{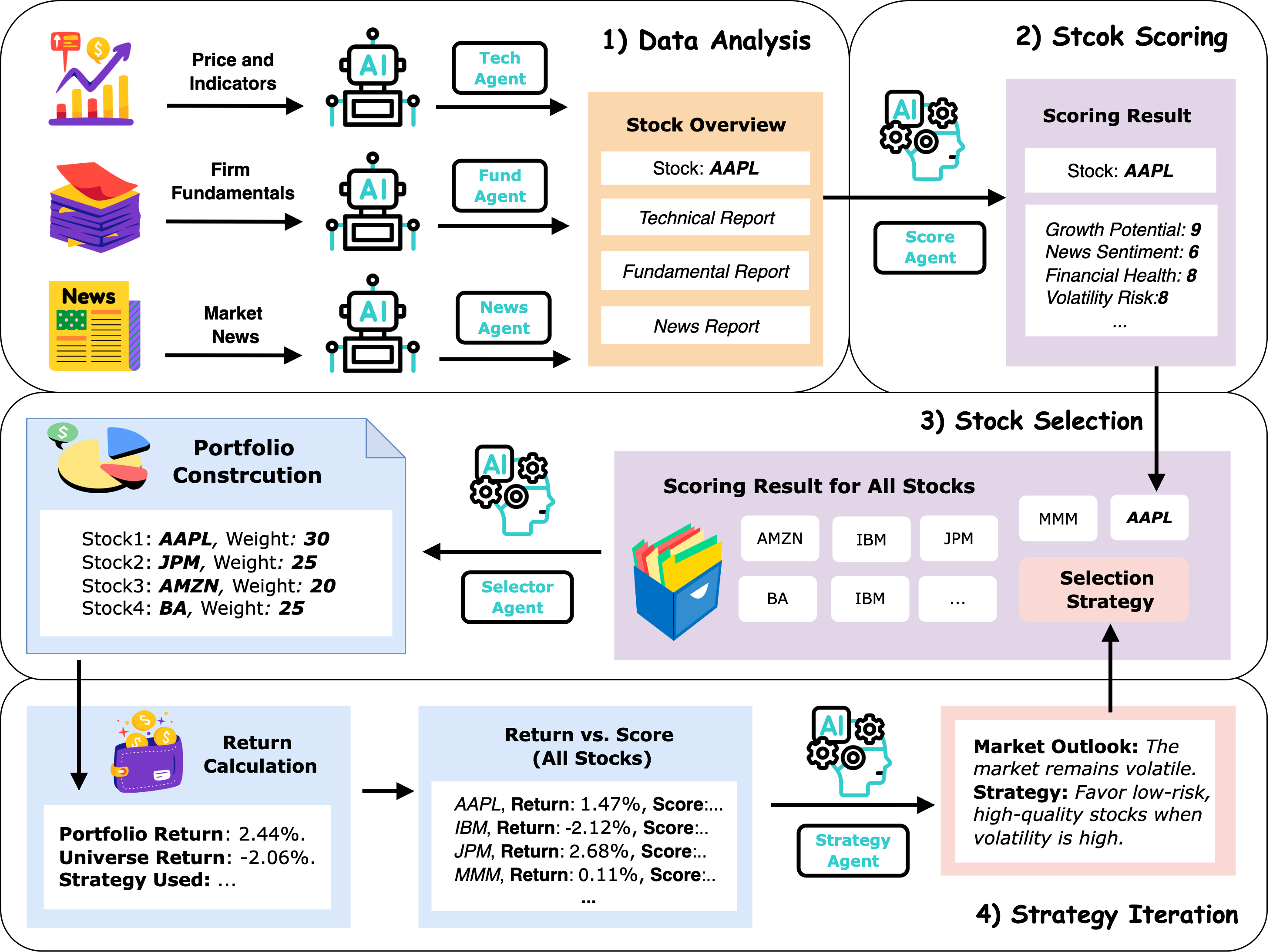}
\caption{3S-Trader Framework.}
\Description{A schematic diagram showing the components of our proposed framework, including scoring, strategy, and selection modules.}
\label{model}
\end{figure*}

\section{Introduction}
In the stock market, a portfolio refers to the structured allocation of various stock assets~\cite{markowitz1952, fabozzi2007robust}, carefully selected to achieve specific investment objectives. Compared to single-stock trading, portfolio-based strategies diversify exposure across multiple assets, mitigating the impact of individual stock volatility~\cite{statman1987diversification}. Constructing an effective portfolio requires analyzing diverse sources of market information, such as price trends, company fundamentals, and macroeconomic signals, to identify the most valuable assets for investment allocation~\cite{grinold2000active, liu2023fingpt}. However, this task becomes increasingly challenging as the number of candidate stocks grows, especially when dealing with large volumes of heterogeneous, multi-modal data that must be processed and compared in a coherent manner~\cite{yang2020deep, kim2024financial}.

With the rapid advancements in AI, particularly LLMs \cite{haque2022think, ahmad2025future}, across domains such as cybersecurity \cite{chopra2024chatnvd, ahmad2025survey, ullah2025skills, goel2025co}, software engineering \cite{goel2024machine, abdulsatar2025towards}, and cloud computing \cite{ahmad2024smart, ahmad2025towards, ahmad2025resilient, jayalath2024microservice}, the technology has also demonstrated significant benefits in the financial sector \cite{Zhang2025regime}. Equipped with powerful natural language understanding and reasoning capabilities \cite{brown2020language, achiam2023gpt, bommasani2021opportunities}, LLMs can analyze and summarize both financial texts \cite{wu2023ptf} and indicators in a way similar to that of human experts and incorporate this information into their trading strategies \cite{kim2024financial, liu2023fingpt}. While the mainstream applications of LLMs in finance have primarily focused on single-stock operations, including price prediction, trend classification (up or down), and position adjustment \cite{gu2024adaptive}. Although such applications can offer useful signals, they are often insufficient for guiding actual stock selection for portfolio management. For instance, a model may predict high returns for a particular stock, but fail to account for associated volatility, making it a less desirable choice in practice. Similarly, strategies that recommend buying or selling individual stocks are not capable of allocating capital across multiple stocks. Furthermore, conventional trading models struggle to learn from past behavior and adapt their strategies accordingly. While reinforcement learning (RL) offers a way to guide output optimization \cite{jiang2017deep}, it often incurs substantial training costs and heavily relies on carefully crafted reward functions \cite{li2023adaptive,deng2016deep}. However, due to the highly volatile nature of financial markets, it becomes challenging for RL-based models to learn robust and generalizable strategies \cite{goel2024optimizing, goel2023enhancing}.

To tackle the challenges outlined above, we propose 3S-Trader, a training-free framework capable of constructing portfolios directly from candidate stocks' recent market signals. Moreover, it can iteratively refine its selection strategy by reflecting on past trading decisions, as illustrated in Figure~\ref{model}.
The framework consists of four main stages. First, in the Data Analysis stage, three specialized LLM-based agents, the News Agent, Fundamental Agent, and Technical Agent, analyze recent market news, company fundamentals, and technical indicators for each stock to generate textual summaries. Next, the Score Agent evaluates each stock along multiple scoring dimensions such as growth potential, volatility risk, and news sentiment, producing stock-level scoring reports. Based on these reports and the current selection strategy (e.g., favoring financially healthy and low-volatility stocks), the Selector Agent selects a subset of stocks and allocates weights to construct the portfolio. Finally, after each trading round, the Strategy Agent analyzes the relationship between candidate scores and their realized returns to identify which types of stocks the market favors, and adjusts its strategy accordingly. Our major contributions are summarized as follows:
\begin{itemize}
    \item We propose a training-free and easily deployable trading framework that directly constructs stock portfolios based on recent market information.
    
    \item We enhance traditional LLM-based trading pipelines by coupling strategy refinement with multi-dimensional scoring, thereby offering trading strategies a clear adjustment direction, without the need for supervision or rewards.
    
    \item We demonstrate the effectiveness and generality of our framework through experiments on stocks from multiple industry sectors, showing its practical value for real-world investment scenarios.
\end{itemize}

\section{Related Work}
\label{sec:related_work}

\textbf{Time-Series Models for Stock Prediction.}
A common paradigm in traditional quantitative finance is to model stock trading as a time series forecasting problem. Auto-regressive models such as ARIMA and GARCH\cite{box2015time, bollerslev1986generalized} are widely adopted for modeling short-term trends and volatility in stock prices. More recently, deep learning techniques have been introduced to model complex temporal dependencies. Long Short-Term Memory (LSTM) networks and Temporal Convolutional Networks (TCNs) have shown promising results in capturing sequential patterns in financial time series \cite{fischer2018deep, borovykh2017conditional}. Transformer-based models like Informer and Stockformer further improve long-range forecasting ability by leveraging attention mechanisms \cite{zhou2021informer,wu2022stockformer}. While these models achieve promising results in forecasting, they primarily focus on price predictions. As a result, they often overlook portfolio-level considerations such as risk control and cross-asset comparison.

\textbf{Multi-LLM Framework for Investment Guidance.}
LLMs have been proposed to leverage their text processing capabilities in financial tasks such as sentiment analysis, financial question answering, and trend prediction. \cite{FinBERT2019, srivastava2024evaluating, koa2024learning}. To achieve a complete pipeline covering the process from information processing to decision generation, frameworks that integrate LLMs with distinct functionalities have been proposed. We categorize these frameworks into two types: \textit{Single-Step} and \textit{Reflective}. The \textit{Single-Step} framework typically aggregates and restructures financial texts over a given period, after which one or more agents summarize the information. A designated output agent then directly relies on this summarized text to generate trading guidance \cite{wu2023finagent, tradingagent2023, liu2023fingpt}. Extending this design, the \textit{Reflective} framework introduces an additional agent to analyze the model’s recent behavior. The analysis is subsequently fed into the next trading cycle, thereby guiding its decision-making process \cite{li2024cryptotrade, liu2024finvision}.

Building on these works, we upgrade the multi-LLM framework by introducing a multi-dimensional scoring mechanism that simplifies summarized information and enables effective cross-stock comparison. Furthermore, by recording historical strategy trajectories and analyzing the relationship between scores and returns, our framework provides more interpretable and iterative strategy refinement.

\section{Preliminaries}
\subsection{Problem Formulation}
\label{sec:Problem}
We consider a weekly stock portfolio construction problem. At the beginning of each week \( t \), a set of stocks \( \mathcal{X} = \{x_1, x_2, \dots, x_n\} \) is available for selection. The goal is to construct a portfolio \( \mathbf{w}_t = [w_t^{(1)}, w_t^{(2)}, \dots, w_t^{(n)}]^\top \), where \( w_t^{(i)} \in [0,1] \) denotes the proportion of capital allocated to stock \( x_i\). To allow for holding cash, we impose the constraint \( \sum_{i=1}^{n} w_t^{(i)} \leq 1 \). 

Trades are executed in the following manner: on the first trading day of week \( t \), capital is allocated to selected stocks according to \( \mathbf{w}_t \) at the opening price; all positions are then liquidated at the closing price on the last trading day of the same week. For each stock \( x_i \in \mathcal{X}\), the weekly return \( r_t^{(i)} \) is calculated as \( r_t^{(i)} = (p_{\text{sell}}^{(i)} - p_{\text{buy}}^{(i)}) / p_{\text{buy}}^{(i)} \), where \( p_{\text{buy}}^{(i)} \) and \( p_{\text{sell}}^{(i)} \) denote the buy and sell prices of stock \( x_i \) in week \( t \), respectively. Given the return vector \( \mathbf{r}_t = [r_t^{(1)}, r_t^{(2)}, \dots, r_t^{(n)}]^\top \), the total portfolio return is computed as \( R_t = \mathbf{w}_t^\top \mathbf{r}_t \).

\subsection{Data Collection}
\label{sec:Data}
We collect three types of data for each stock: stock price and technical indicators, firm fundamentals, and market news. In this subsection, we describe how each type of data is processed for the portfolio construction task at week \( t \).

\subsubsection{Price and Technical Indicators}
The raw price data consists of daily stock prices and trading volume. Based on these time series, we compute several technical indicators, including SMA, ATR, RSI, MACD, and Bollinger Bands. These indicators are commonly used in financial technical analysis~\cite{murphy1999technical,ta2019library}.

At week \( t \), we extract the daily closing prices \( \text{Price}_{i,d} \) and technical indicators \( \text{Indicators}_{i,d} \) for stock \( x_i \) over the preceding four calendar weeks \( \mathcal{W}_{t-4:t-1} \), where \( d \) indexes the days within that range. These values are concatenated into a plain text \( \text{tech}_{i,t} \), denoted as the technical input of stock \( x_i \) for week \( t \):
\begin{equation}
    \text{tech}_{i,t} = \text{ConcatText}(\{\text{Price}_{i,d}, \text{Indicators}_{i,d} \mid d \in \mathcal{W}_{t-4:t-1}\})
    \label{eq:tech_input}
\end{equation}

\subsubsection{Market News} 
The raw news data includes article titles, summaries, and their associated stock symbols. At week \( t \), we collect all news items related to stock \( x_i \) from the previous week. Similar to Equation~\eqref{eq:tech_input}, we define the news input as:
\begin{equation}
    \text{news}_{i,t} = \text{ConcatText}(\{\text{RawNews}_{i,d} \mid d \in \mathcal{W}_{t-1}\})
    \label{eq:news_input}
\end{equation}
Here, \( \text{RawNews}_{i,d} \) denotes the news content related to stock \( x_i \) published on day \( d \), and \( \mathcal{W}_{t-1} \) represents the calendar week immediately preceding week \( t \).

\subsubsection{Firm Fundamentals} 
The fundamental data consists of firm-specific earnings reports, balance sheets, and cash flow statements, which are updated on a quarterly basis. To enable trend analysis, we concatenate the fundamental records from the four most recent fiscal quarters released before week \( t \). The resulting text serves as the fundamental input of stock \( x_i \) at week \( t \), denoted as:
\begin{equation}
    \text{fund}_{i,t} = \text{ConcatText}(\{\text{RawFund}_{i,q} \mid q \in \mathcal{Q}_{t_q-3:t_q}\})
    \label{eq:fund_input}
\end{equation}
Here, \( \text{RawFund}_{i,q} \) represents the fundamental data of stock \( x_i \) reported in fiscal quarter \( q \), and \( \mathcal{Q}_{t_q-3:t_q} \) refers to the four most recent quarters available before week \( t \).

\section{Methodology}
In this section, we introduce the architecture of our proposed framework, 3S-Trader. As shown in Figure~\ref{model}, the overall workflow is structured into four main steps. The framework incorporates six LLM-based agents that collaborate across these steps to support adaptive portfolio management. We implement all LLM-based agents using GPT-4o\footnote{\url{https://openai.com/gpt-4o}}. In the following subsections, we detail each step and explain how the agents interact to process information, refine strategies, and make investment decisions.

\subsection{Market Analysis}
\begin{figure}[h]
    \centering
    \includegraphics[width=0.9\linewidth]{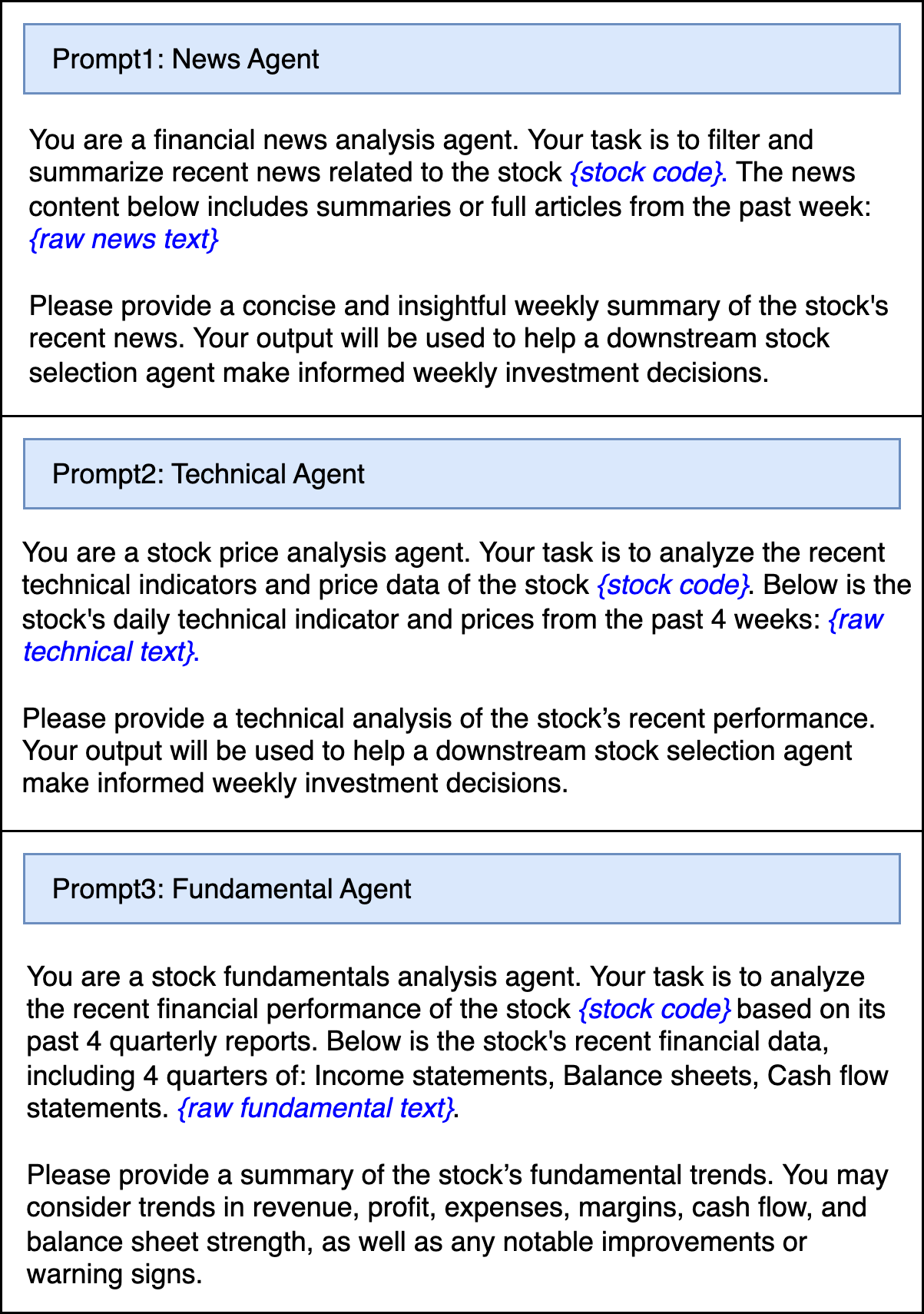}
    \caption{Example prompts used by News Agent, Technical Agent, and Fundamental Agent for market analysis.}
    \label{fig:prompt_analysis}
\end{figure}

In this part, we leverage three specialized agents: \textit{News Agent}, \textit{Fundamental Agent}, and \textit{Technical Agent}, to analyze different types of input data. Specifically, at week \( t \), for each stock \( x_i \), based on its aggregated news text \( \text{news}_{i,t} \) defined in Equation~\eqref{eq:news_input}, the \textit{News Agent} generates an analysis report $\alpha^{\text{news}}_{i,t}$ as follows:
\begin{equation}
    \alpha^{\text{news}}_{i,t} = \text{agent}_\text{news}(\text{news}_{i,t}, \text{prompt}_{\text{news}})
\end{equation}
Similarly, we obtain $\alpha^{\text{tech}}_{i,t}$ and $\alpha^{\text{fund}}_{i,t}$ from the \textit{Technical Agent} and \textit{Fundamental Agent}, respectively, using corresponding inputs and prompts. These outputs are then concatenated to form the full data overview of stock \( x_i \) at week \( t \), denoted as \( o_{i,t} \):
\begin{equation}
    o_{i,t} = \text{ConcatText}(\alpha^{\text{news}}_{i,t}, \alpha^{\text{tech}}_{i,t}, \alpha^{\text{fund}}_{i,t})
    \label{eq:overview}
\end{equation}
Example prompts used by the corresponding agents are illustrated in Figure~\ref{fig:prompt_analysis}.

\subsection{Stock Scoring}

In this part, the \textit{Score Agent} performs multi-dimensional scoring based on each stock’s recent market signals. It is instructed to assign a score from 1 to 10 for each dimension, reflecting the stock’s relative strength in that aspect.

In designing the scoring dimensions, we focused on their relevance to the available input data. To ensure that each dimension is grounded in observable evidence, we derive them directly from three primary sources: stock price movements, firm fundamentals, and market news. Based on these, we define the following six dimensions:

\begin{itemize}
    \item \textbf{Financial Health}: Evaluates a company’s current financial stability. A higher score reflects stronger fundamentals and lower short-term risk.

    \item \textbf{Growth Potential}: Assesses the company's future expansion capacity based on investment plans, and industry growth outlook. A higher score suggests stronger long-term earnings potential.

    \item \textbf{News Sentiment}: Reflects overall sentiment polarity extracted from recent news articles. A higher score implies more positive news coverage and investor perception.

    \item \textbf{News Impact}: Assesses the breadth and duration of news influence. Higher scores reflect more sustained impacts, e.g., from political events or industry-level shifts.

    \item \textbf{Price Momentum}: Captures recent upward or downward trends in stock price movement. A higher score reflects a stronger and more consistent upward price trend.

    \item \textbf{Volatility Risk}: Quantifies the level of recent price fluctuations, indicating risk exposure. A higher score represents higher volatility and less stable price behavior.
\end{itemize}

For stock selection at week \( t \), we first obtain the data overview \( o_{i,t} \) for stock \( x_i \) as defined in Equation~\eqref{eq:overview}. This overview is then processed by the Score Agent to produce a textual scoring result, denoted as \( s_{i,t} \). This scoring process can be formalized as:
\begin{equation}
    s_{i,t} = \text{agent}_{\text{score}}(o_{i,t}, \text{prompt}_{\text{score}})
\end{equation}

\begin{figure}[t]
    \centering
    \includegraphics[width=0.9\linewidth]{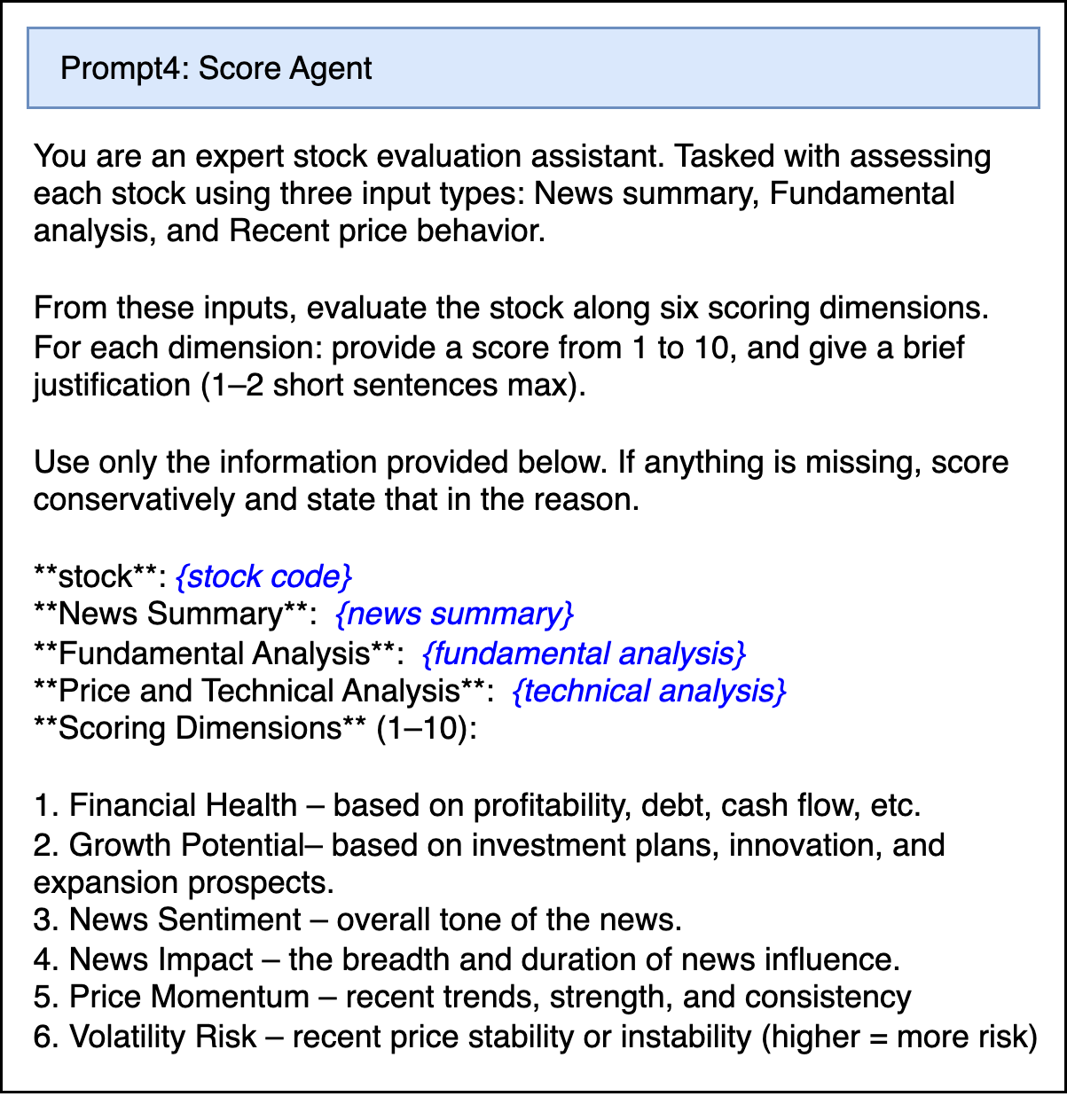}
    \caption{Example prompt used by the Score Agent for stock scoring.}
    \label{fig:prompt_score}
\end{figure}

An example prompt used by the Score Agent is illustrated in Figure~\ref{fig:prompt_score}.

\subsection{Stock Selection}

In this part, the \textit{Selector Agent} constructs the final portfolio based on two key inputs: the scoring results of all candidate stocks produced by the \textit{Score Agent}, and the selection strategy provided by the \textit{Strategy Agent}. For clarity, we will introduce the \textit{Strategy Agent} later. Here, it suffices to note that the strategy is a textual description \( \pi_t \) that specifies which types of stocks should be preferred, guiding the \textit{Selector Agent} to prioritize stocks with higher scores in the relevant dimensions. For example, a strategy \( \pi_t \) might be:

\textit{Increase emphasis on financial health and reduce exposure to high-volatility stocks, as recent returns indicate stronger performance from fundamentally stable companies.}

\begin{figure}[h]
    \centering
    \includegraphics[width=0.9\linewidth]{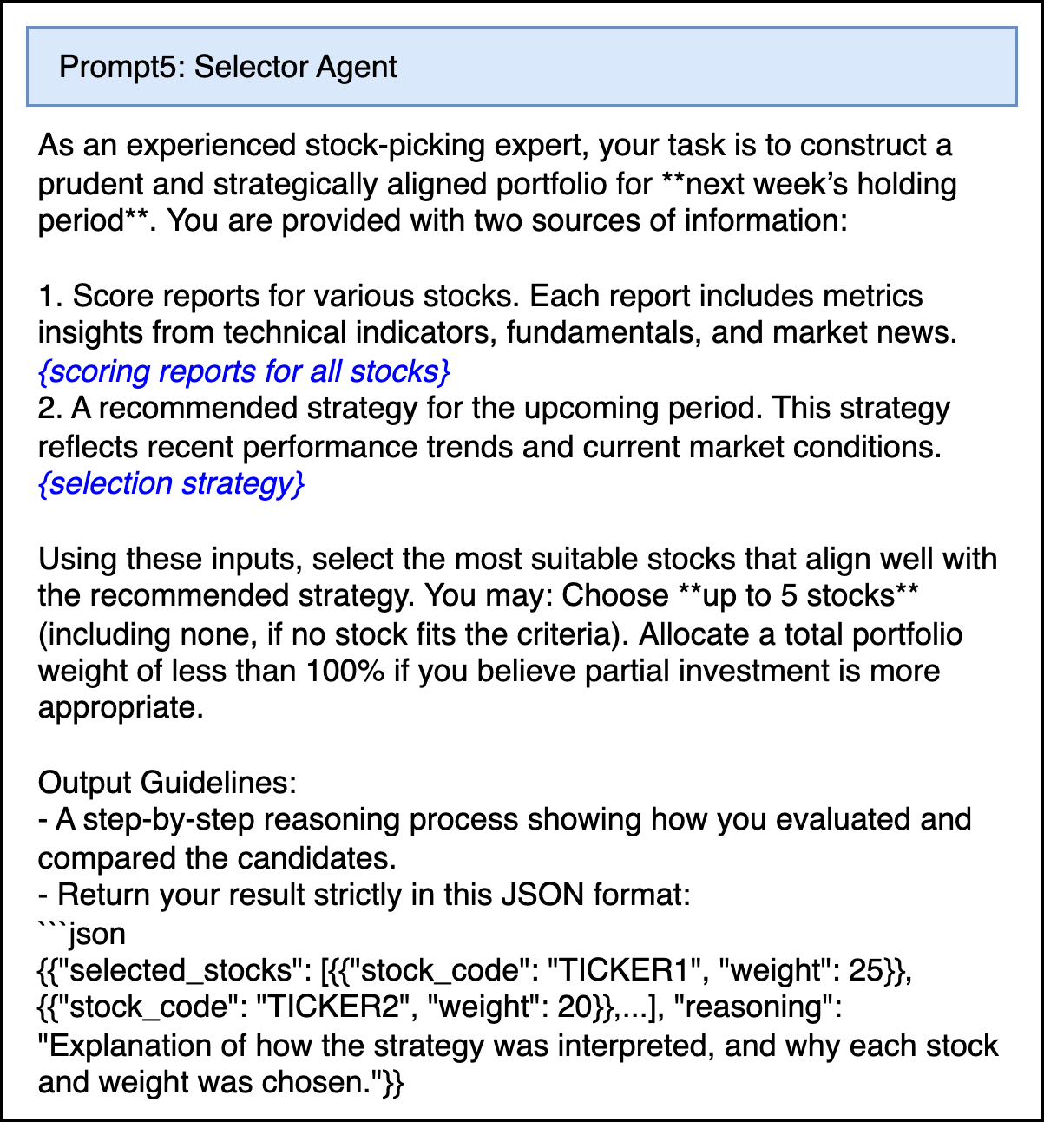}
    \caption{Example prompt used by the Selector Agent for stock selection.}
    \label{fig:prompt_selector}
\end{figure}

The \textit{Selector Agent} is tasked with selecting up to five stocks and assigning weights, allowing cash holding to avoid market downturns. Hence, the total portfolio weight may be less than 1. For the portfolio construction task at week \( t \), the agent receives the scores \( s_{i,t} \) for all candidate stocks \( x_i \in \mathcal{X} \), along with the strategy \( \pi_t \). The resulting portfolio can be expressed as:
\begin{equation}
    \mathbf{w}_t = \text{agent}_{\text{select}}(\text{ConcatText}(\{s_{i,t}\}_{i=1}^n), \pi_t, \text{prompt}_{\text{select}})
    \label{eq:selector_output}
\end{equation}

where \( \mathbf{w}_t = [w_t^{(1)}, w_t^{(2)}, \dots, w_t^{(n)}]^\top \) is the portfolio weight vector, with \( \sum_{i=1}^n w_t^{(i)} \leq 1 \), and at most 5 elements of \( \mathbf{w}_t \) are nonzero. An example prompt used by the Selector Agent is illustrated in Figure~\ref{fig:prompt_selector}.

\subsection{Strategy Iteration}

In this part, the \textit{Strategy Agent} refines the current strategy based on the realized price changes of each stock and the historical trajectory of past strategies.

At the end of week \( t \), following the trading rule defined in Section~\ref{sec:Problem}, we obtain the realized return \( r_t^{(i)} \) for each stock \( x_i \). To support strategic refinement, the \textit{Strategy Agent} receives the realized returns \( r_t^{(i)} \) of all candidate stocks and their corresponding scoring reports \( s_{i,t} \). It is tasked with identifying shared characteristics among high- and low-performing stocks, and proposing updated strategies.

\begin{figure}[t]
    \centering
    \includegraphics[width=0.9\linewidth]{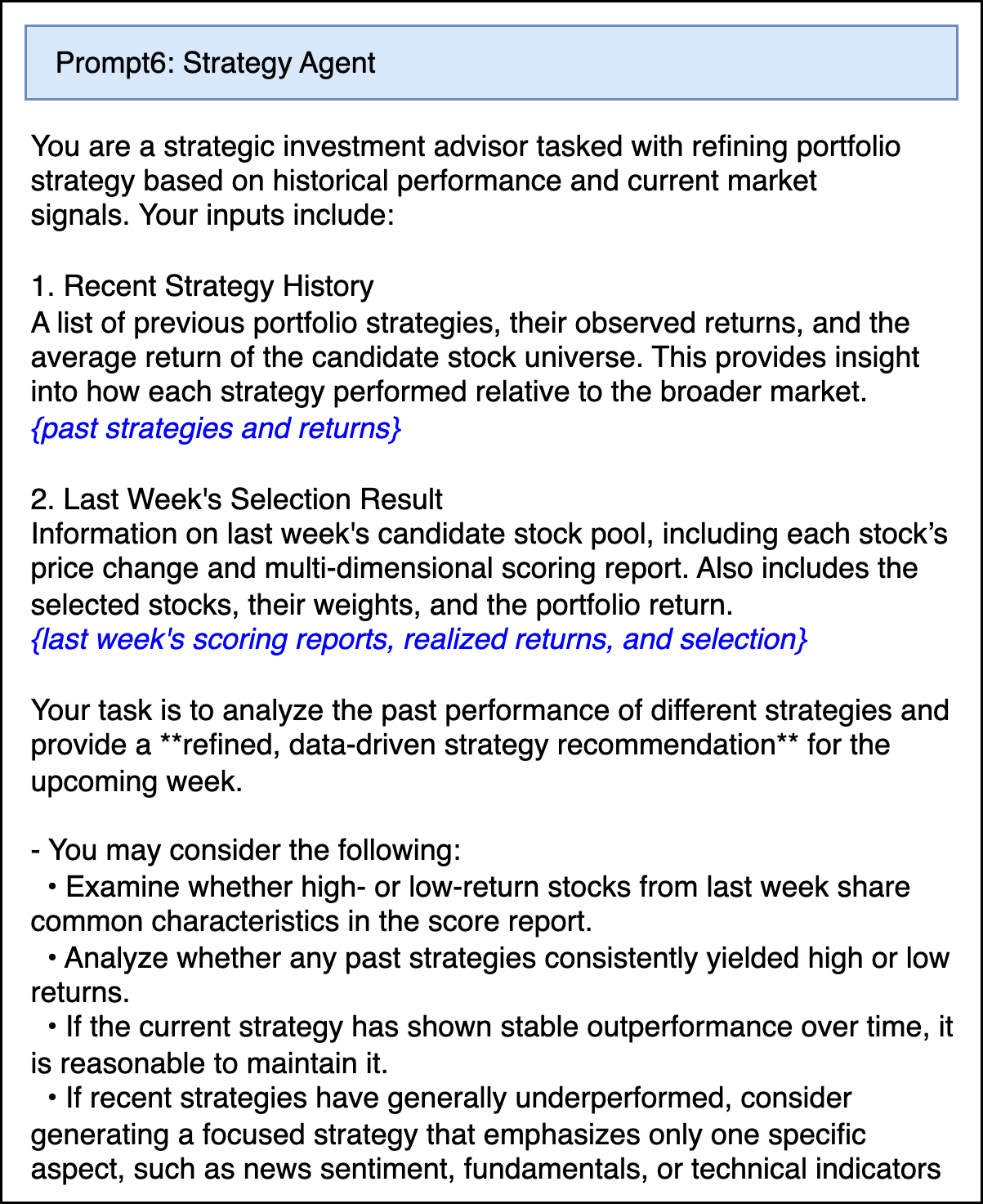}
    \caption{Example prompt used by the Strategy Agent for strategy iteration.}
    \label{fig:prompt_strategy}
\end{figure}

To ensure a stable and coherent strategy iteration process, we introduce the trajectory of past strategy updates into the input of the \textit{Strategy Agent}. This design aims to prevent divergence from previously effective strategies, reduce unnecessary fluctuations in decision-making, and help the agent identify patterns that are indicative of long-term stable returns. The historical trajectory is denoted as:
\begin{equation}
\mathcal{H}t = \left\{ \text{ConcatText} \left( \pi_{t-k}, R_{t-k}^{\text{avg}}, R_{t-k} \right) \right\}_{k=1}^K
\end{equation}
where \( \pi_{t-k} \) denotes the strategy adopted in week \( t-k \),  
\( R_{t-k}^{\text{avg}} \) is the universe average return,  
and \( R_{t-k} \) is the portfolio return. We set K = 10 in our experiments, enabling the review of the previous 10 weeks. Subsequently, the refined strategy for the next week can be defined as follows:
\begin{equation}
    \pi_{t+1} = \text{agent}_{\text{strategy}}(\pi_{t}, \mathbf{w}_{t}, \mathbf{r}_{t}, \mathbf{s}_{t}, \mathcal{H}_t, \text{prompt}_{\text{strategy}})
    \label{eq:strategy_update}
\end{equation}
where \( \mathbf{w}_t \) is the portfolio weight vector selected in week \( t \),  
\( \mathbf{r}_t = [r_t^{(1)}, r_t^{(2)}, \dots, r_t^{(n)}]^\top \) is the realized return vector of all candidate stocks,  
and \( \mathbf{s}_{t} = \{s_{i,t}\}_{i=1}^n \) denotes the set of scoring reports for all stocks in week \( t \).  
An example prompt used by the Strategy Agent is illustrated in Figure~\ref{fig:prompt_strategy}.

\section{Experiments}
In this section, we detail our experimental setup and evaluate the proposed framework by conducting portfolio construction across various stock universes. The performance of our framework is compared against several baseline models to demonstrate its effectiveness and applicability.

\subsection{Experimental Settings}
\subsubsection{Dataset Description}
We construct four distinct candidate stock universes for evaluation. For each universe, we independently conduct experiments to assess the model's performance under varying market conditions.

\begin{itemize}
    \item \textbf{DJIA Constituents}: This set includes 30 large-cap U.S. companies from diverse sectors and serves as a representative benchmark of the broader market\footnote{\url{https://en.wikipedia.org/wiki/Dow_Jones_Industrial_Average}}.
    
    \item \textbf{Technology Sector Stocks}: Comprising 44 constituent companies from the NASDAQ-100 Technology Sector Index\footnote{\url{https://indexes.nasdaqomx.com/Index/Weighting/NDXT}}.
    
    \item \textbf{Financial Sector Stocks}: A subset of 49 major companies with the highest weights in the SPDR Fund for the Financial Select Sector\footnote{\url{https://www.ssga.com/us/en/individual/etfs/the-financial-select-sector-spdr-fund-xlf}}. 
    
    \item \textbf{Healthcare Sector Stocks}: A selection of 46 top-weighted companies from the SPDR Fund for the Select Sector of Health Care\footnote{\url{https://www.ssga.com/us/en/individual/etfs/the-health-care-select-sector-spdr-fund-xlv}}.
\end{itemize}

Data are retrieved from Alpha Vantage \footnote{\url{https://www.alphavantage.co/}}, and preprocessed according to the methodology detailed in Section~\ref{sec:Data}. The evaluation period spans from May 16, 2022 to May 27, 2024. For time-series baselines that require training, we provide historical price data covering the period from May 1, 2012 to May 15, 2022 as the training and validation set.

\subsubsection{Baselines}
We use the equal-weight strategy (1/N) as a market reference and compare our method against three representative categories of models.

\begin{itemize}
\item \textbf{Rule-based methods:} Including SMA (Simple Moving Average), MACD (Moving Average Convergence Divergence), and BOLL (Bollinger Bands) \cite{murphy1999technical}. For each indicator, we construct a corresponding factor score and select the top 5 stocks each week, assigning 20\% weight to each.

\item \textbf{Deep learning prediction models:} We include the classical LSTM model~\cite{hochreiter1997long} as well as two state-of-the-art transformer-based models: Informer~\cite{zhou2021informer} and Autoformer~\cite{wu2021autoformer}. For each stock \( i \) at week \( t \), the response variable is the weekly return \( r_{i,t} \). The input to the model includes a 4-week time window of technical features, combined with a stock-specific embedding to enable multi-stock prediction. All model hyper-parameters are tuned on a validation set. At each week \( t \), the top five stocks with the highest predicted returns are assigned an equal weight of 20\% each.

\item \textbf{Multi-LLM baselines:} We include both a single-step version and a reflective variant as described in Section~\ref{sec:related_work}. The single-step framework, implemented following by TradingAgent~\cite{tradingagent2023}, directly generates portfolios from summarized information. The reflective version, following the design of CryptoTrade~\cite{li2024cryptotrade}, further incorporates a \textit{Reflection Agent} that analyzes previous portfolio returns to refine stock selection. For consistency and fair comparison, we make minor adjustments to the input and output structures of both implementations.
\end{itemize}

\subsubsection{Evaluation Metrics}
We use three metrics to evaluate portfolio performance: Accumulated Return (AR), 
Sharpe Ratio (SR), and Calmar Ratio (CR). 

\begin{itemize}
\item \textbf{Accumulated Return (AR)}: AR is defined as the total compounded return over the evaluation period:
\begin{equation}
\text{AR} = \prod_{t=1}^T (1 + R_t) - 1
\label{eq:ar}
\end{equation}
where \( R_t \) denotes the portfolio return at week \( t \), and \( T \) is the total number of evaluation weeks.

\item \textbf{Sharpe Ratio (SR)}: SR measures the risk-adjusted return, capturing how efficiently a portfolio converts volatility into excess return. It is calculated as:
\begin{equation}
\text{SR} = \frac{\mathbb{E}[R_t]}{\sigma(R_t)}
\label{eq:sr}
\end{equation}
assuming a zero risk-free rate. Here, \( \mathbb{E}[R_t] \) is the mean return and \( \sigma(R_t) \) is the standard deviation of returns.

\item \textbf{Calmar Ratio (CR)}: CR evaluates return relative to the worst drawdown observed, providing an indication of how well a strategy balances profitability against downside risk. It is defined as:
\begin{equation}
\text{CR} = \frac{\text{AR}}{|\text{MDD}|}
\label{eq:cr}
\end{equation}
where the maximum drawdown (MDD) is computed as:
\begin{equation}
\text{MDD} = \min_t \left( \frac{C_t - \max_{i \le t} C_i}{\max_{i \le t} C_i} \right)
\label{eq:mdd}
\end{equation}
and \( C_t \) is the accumulated return up to week \( t \).
\end{itemize}

\subsection{Results and Analysis}

We summarize the experimental results across different stock universes in a single table, as shown in Table~\ref{tab:performance_comparison}. Below, we present our key findings.

\begin{table*}[h]
\centering
\caption{Performance comparison across different models and stock universes. \textcolor{red}{Red} indicates the highest value in each column, and \textcolor{blue}{blue} indicates the second highest.}
\label{tab:performance_comparison}
\begin{tabular}{llccccccccccccc}
\toprule
\multirow{2}{*}{Category} & \multirow{2}{*}{Approach} 
& \multicolumn{3}{c}{DJIA Constituents} 
& \multicolumn{3}{c}{Financial Sector} 
& \multicolumn{3}{c}{Healthcare Sector} 
& \multicolumn{3}{c}{Technology Sector} \\
\cmidrule(r){3-5} \cmidrule(r){6-8} \cmidrule(r){9-11} \cmidrule(r){12-14}
& & AR(\%) & SR & CR & AR(\%) & SR & CR & AR(\%) & SR & CR & AR(\%) & SR & CR \\
\midrule
\multirow{1}{*}{Benchmark} 
& 1/N                          & 39.50   & 0.168 & 2.98 & 52.63  & 0.16 & 3.95 & 27.82  & 0.12 & 2.45 & 72.13  & 0.16 & 3.59 \\
\midrule
\multirow{3}{*}{Rule-based}
& SMA                          & 70.63   & 0.20 & 4.52 & 33.10  & 0.10 & 2.08 & 17.21  & 0.07 & 1.26 & 67.36  & 0.13 & 1.95 \\
& MACD                         & 46.00   & 0.15 & 3.02 & 61.16  & 0.15 & 3.42 & 20.58  & 0.08 & 1.29 & 116.29 & 0.19 & 4.30 \\
& BOLL                         & 20.09   & 0.08 & 1.20 & 25.09  & 0.06 & 1.60 & 6.82   & 0.04 & 0.51 & 96.98  & 0.16 & 5.16 \\
\midrule
\multirow{3}{*}{Deep Learning}
& LSTM                         & 88.42  & 0.23 & 4.91 & 61.24  & 0.15 & 3.54 & \textcolor{red}{59.78}  & 0.15 & 3.26 & \textcolor{red}{193.39} & 0.21 & 5.81 \\
& Informer                     & 68.96   & 0.22 & 3.88 & \textcolor{red}{89.49}  & \textcolor{red}{0.25} & \textcolor{red}{8.08} & 44.37  & 0.15 & 3.54 & 98.61  & 0.15 & 3.33 \\
& Autoformer                   & \textcolor{blue}{104.26}  & 0.24 & 5.35 & 75.92  & 0.17 & 4.50 & 30.44  & 0.11 & 2.30 & 102.90 & 0.14 & 2.97 \\
\midrule
\multirow{3}{*}{Multi-LLM}
& Single-Step       & 91.69  & 0.27 & 8.94 & 34.92  & 0.15 & 3.10 & 31.40  & 0.12 & 3.11 & 152.84 & \textcolor{blue}{0.25} & \textcolor{blue}{9.04} \\
& Reflective         & 96.61  & \textcolor{blue}{0.28} & \textcolor{blue}{9.90} & 44.86  & 0.17 & 4.56 & 43.73  & \textcolor{red}{0.18} & \textcolor{red}{6.51} & 128.27 & 0.21 & 5.41 \\
& \textbf{3S-Trader} & \textcolor{red}{131.83} & \textcolor{red}{0.31} & \textcolor{red}{11.84} & \textcolor{blue}{84.93} & \textcolor{blue}{0.21} & \textcolor{blue}{7.57} & \textcolor{blue}{51.41} & \textcolor{blue}{0.17} & \textcolor{blue}{3.82} & \textcolor{blue}{183.29} & \textcolor{red}{0.27} & \textcolor{red}{11.81} \\
\bottomrule
\end{tabular}
\end{table*}

\subsubsection{Overall Performance of 3S-Trader}
\begin{figure*}[h]
    \centering
    \includegraphics[width=0.9\linewidth]{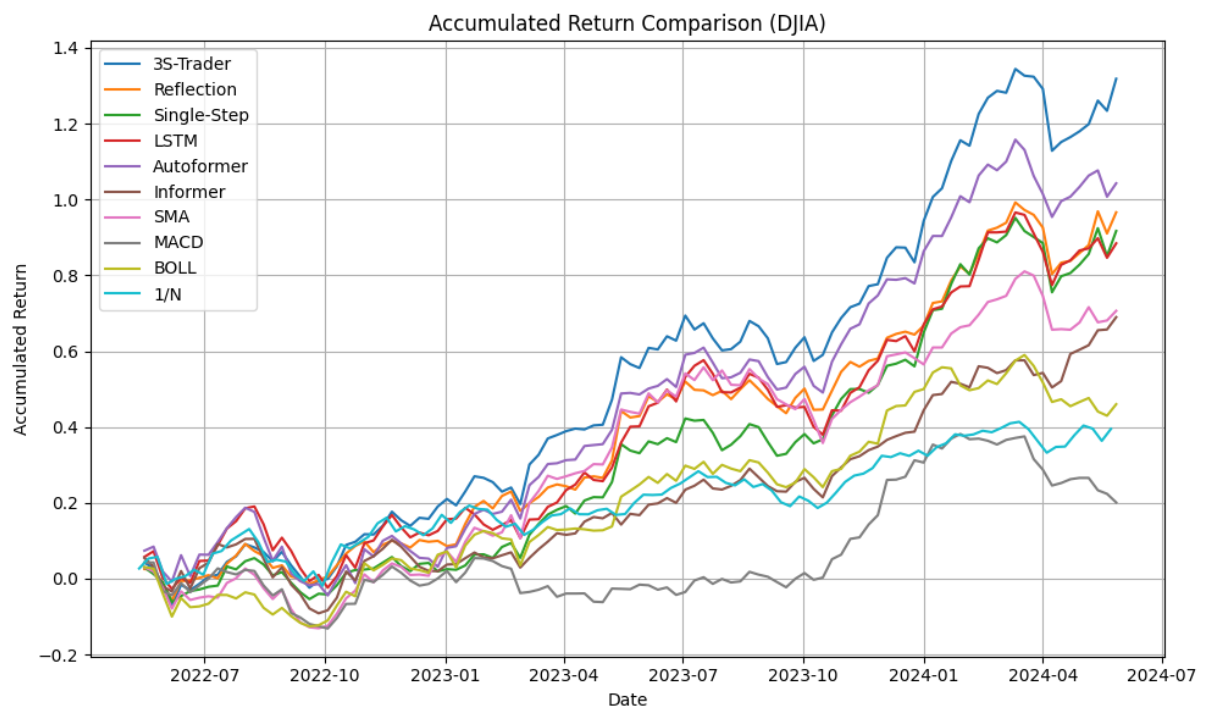}
    \caption{Comparison of accumulated returns for all methods on DJIA constituents.}
    \label{fig:djia_comparison}
\end{figure*}

Our proposed method demonstrates stable performance across diverse market environments, consistently ranking among the top two models across nearly all evaluation metrics. This advantage is most evident in the mixed-sector DJIA constituents, where 3S-Trader achieves the highest accumulated return of 131.83\%, significantly outperforming the second-best model. As illustrated in Figure~\ref{fig:djia_comparison}, the upward trend of 3S-Trader is clearly visible. Moreover, 3S-Trader exhibits no obvious weaknesses across different performance metrics and sectors, indicating a high degree of robustness and stability in its returns.

\subsubsection{Comparison with Rule-based Models}
Our proposed method significantly outperforms traditional rule-based strategies across all sectors. In many cases, classical technical indicators such as SMA, MACD, and BOLL even underperform the simple 1/N benchmark. This underperformance can be attributed, in part, to the experimental design: to ensure fair comparison and interpretability, we use fixed-factor rule-based strategies with relatively low trading frequency. However, in real-world trading scenarios, the effectiveness of such strategies often depends on dynamic factor selection, regular parameter tuning, and more frequent position adjustments. These complexities are difficult to fully capture within our current experimental framework. By contrast, our LLM-based approach adapts more flexibly to changing market conditions and integrates diverse signals in a coherent and data-driven manner, which explains its consistent advantage over static rule-based methods.

\subsubsection{Comparison with Deep Learning Models}
Deep learning models demonstrate strong performance in terms of accumulated return, with each of the three sector-specific stock pools featuring a deep learning model as the top performer in this dimension. In particular, the classical sequence prediction model LSTM performs competitively against the more recent state-of-the-art architectures like Informer and Autoformer. In the technology sector, for instance, LSTM achieves an accumulated return of 193.39\%, nearly double that of Informer (98.61\%) and Autoformer (102.90\%), highlighting its capacity to capture strong trends in high-momentum markets.

However, these models exhibit significant limitations in terms of risk control. This is reflected in both the Sharpe and Calmar Ratios, where their performance often falls short compared to LLM-based frameworks. The relatively lower risk-adjusted returns suggest that deep learning models may overfit to past trends or struggle to generalize under market volatility, leading to inconsistent or overly aggressive allocations. In contrast, multi-LLM systems show more balanced performance, with consistently higher Sharpe and Calmar Ratios. This indicates that while deep learning models are good at capturing trends, they often lack the ability to reason and handle complex information. Multi-LLM frameworks, by combining different data sources and using strategy texts, are better at making stable and informed decisions under uncertainty.

\subsubsection{Self-Refined Frameworks in Volatile Markets}
\begin{figure}[t]
    \centering
    \includegraphics[width=0.9\linewidth]{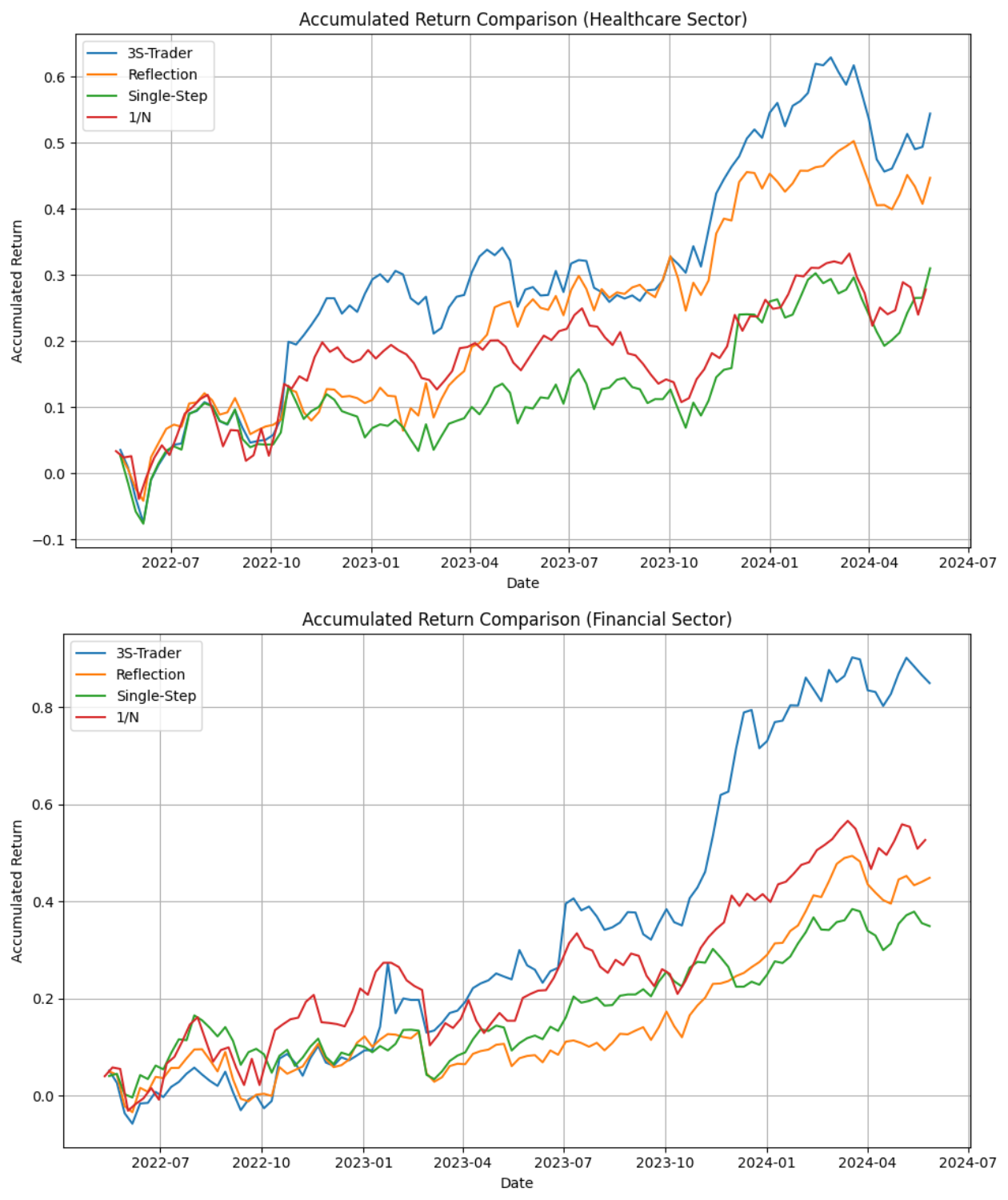}
    \caption{Comparison of accumulated returns for 3S-Trader, Reflective and Single-Step Frameworks across Healthcare and Financial sectors.}
    \label{fig:sector_comparison}
\end{figure}

Both 3S-Trader and the \textit{Reflective} framework are designed to refine their strategies by learning from past decisions. We focus on their performance in volatile markets such as healthcare and finance, as shown in Figure~\ref{fig:sector_comparison}. In the healthcare sector, the performance gap between the two frameworks is not substantial, both demonstrate clear improvements compared to the \textit{Single-Step} baseline. Notably, the \textit{Reflective} variant achieves the best stability in this sector, recording the highest SR of 0.18 and CR of 6.51.

However, in the financial sector, the performance gain from the reflection mechanism is less pronounced. The \textit{Reflective} variant even underperforms both the \textit{Single-Step} baseline and the market benchmark for most of the validation period, potentially due to over-adjustment or the absence of clear directional guidance during strategy updates. In contrast, 3S-Trader not only incorporates a reflective loop but also leverages a multi-dimensional scoring system that provides explicit and interpretable criteria for strategy refinement. This holistic design enables consistently superior and more stable performance across different market conditions.

\section{Conclusion and Future Work}
In this paper, we proposed 3S-Trader, a multi-LLM framework for portfolio construction that is capable of self-adjustment and adapts to diverse market conditions. The framework condenses recent market information into stock-level scoring reports and applies explicit, interpretable selection criteria to guide portfolio allocation. Compared with traditional rule-based and time-series forecasting models, our approach requires no model training or parameter tuning, yet achieves consistently strong performance across different stock universes, delivering competitive returns and robust stability.

For future work, several directions remain open. First, while the current scoring dimensions are designed based on domain expertise, a promising extension is to enable the automatic discovery and learning of scoring factors from data. Second, our experiments are limited to backtesting; validating the framework in live trading environments will be essential to assess its real-world feasibility. Lastly, incorporating broader asset classes and exploring dynamic risk-control mechanisms could further enhance the generalizability and practical value of the proposed system.

\bibliographystyle{ACM-Reference-Format}
\bibliography{base}

\end{document}